\documentclass[aps,pre,twocolumn,superscriptaddress,nofootinbib,longbibliography]{revtex4-2}

\usepackage{amsmath,amssymb,bm}
\newcommand{\abs}[1]{\left|#1\right|}
\usepackage{graphicx}
\usepackage{hyperref}
\usepackage{siunitx}
\usepackage{orcidlink}

\newcommand{\ii}{\mathrm{i}}
\newcommand{\Tr}{\mathrm{tr}}

\begin{document}

\title{Energetic vs Inference-Based Invisibility: Fisher-Information Analysis of Two-Layer Acoustic Near-Cloaks}

\author{J.\ Sumaya-Martinez\,\orcidlink{0000-0002-7032-8824}}
\affiliation{Facultad de Ciencias, Universidad Autónoma del Estado de México, Toluca, México}

\author{J.\ Mulia-Rodriguez}
\affiliation{Facultad de Ciencias, Universidad Autónoma del Estado de México, Toluca, México}

\date{\today}

\begin{abstract}
Near-cloaks based on passive coatings can strongly suppress the scattered-field energy in a narrow frequency band, yet an observer’s ability to
\emph{infer} object parameters from noisy measurements need not decrease proportionally. We develop a fully theoretical, two-dimensional (2D) framework
for a coated acoustic cylinder in an air background. Using an exact cylindrical-harmonic solution of the Helmholtz equation, we compute the modal
scattering coefficients $a_m(\omega)$ for a core of radius $a$ surrounded by two concentric, effective-fluid layers. We design the coating to cancel the
dominant low-order multipoles (monopole $m=0$ and dipole $m=\pm 1$) at a target frequency, yielding a narrowband near-cloak.
Beyond the conventional energetic metric (total scattering width), we quantify \emph{information-based detectability} through the Fisher information matrix
(FIM) and the associated Cram\'er--Rao lower bounds (CRLBs) for joint estimation of size and material parameters, $\bm{x}=[a,\rho_1,c_1]^T$, from noisy
far-field data. A representative air-background case study exhibits a peak $\sim25$~dB reduction in total scattering width near the design frequency, while
$\Tr(\mathrm{FIM})$ decreases by only a few dB, illustrating that energy-based and inference-based invisibility are distinct objectives.
We discuss bandwidth limitations consistent with passivity/causality and delay--bandwidth constraints, and show how the FIM/CRLB viewpoint provides a
task-aware and parameter-resolved characterization of near-cloaking.
\end{abstract}

\maketitle

\section{Introduction}
Acoustic cloaking aims to mitigate an object's scattering signature, ideally rendering it indistinguishable from the surrounding medium.
Two broad paradigms dominate the literature: transformation acoustics, which prescribes spatially varying anisotropic parameters to steer waves
around a hidden region~\cite{CummerSchurig2007}, and scattering cancellation (multipole cancellation), which suppresses dominant low-order scattering
harmonics using coatings or metasurfaces~\cite{GuildAluHaberman2011}. Transformation-based cloaks are elegant but often require extreme or anisotropic
parameters that are difficult to realize---particularly in air---whereas scattering-cancellation designs are analytically tractable for canonical
geometries and naturally lead to bandwidth-limited \emph{near}-cloaks.

Fundamental constraints further motivate near-cloaks: passivity and causality impose stringent trade-offs between achievable scattering reduction and
bandwidth~\cite{MonticoneAlu2016}, and delay--bandwidth considerations restrict cloaking of electrically/acoustically large objects~\cite{HashemiZhang2010}.
Consequently, a physically meaningful target is frequently near-cloaking---strong suppression in a narrow band---rather than ideal broadband invisibility.

Most cloaking studies quantify performance using energetic metrics, such as total scattering cross section/width. In many detection and classification
settings, however, an observer performs \emph{inference}: estimating object parameters from noisy measurements.
This motivates an information-theoretic viewpoint. We quantify parameter detectability using the Fisher information matrix (FIM) and the associated
Cram\'er--Rao lower bounds (CRLBs), which provide fundamental limits on any unbiased estimator~\cite{GustafssonNordebo2006,NordeboJMP2012}.

\textbf{Contributions.} (i) We present a compact exact modal formulation for a two-layer coated 2D cylinder in an air background. (ii) We design the coating
to suppress the dominant monopole and dipole scattering coefficients at a target frequency, yielding a near-cloak. (iii) We analyze joint size--material
inference for $\bm{x}=[a,\rho_1,c_1]^T$ and demonstrate that energy-based invisibility and FIM/CRLB-based detectability can differ substantially.

\section{Model and exact scattering formulation}
\subsection{Geometry and governing equations}
\begin{figure}[t]
\centering
\includegraphics[width=\columnwidth]{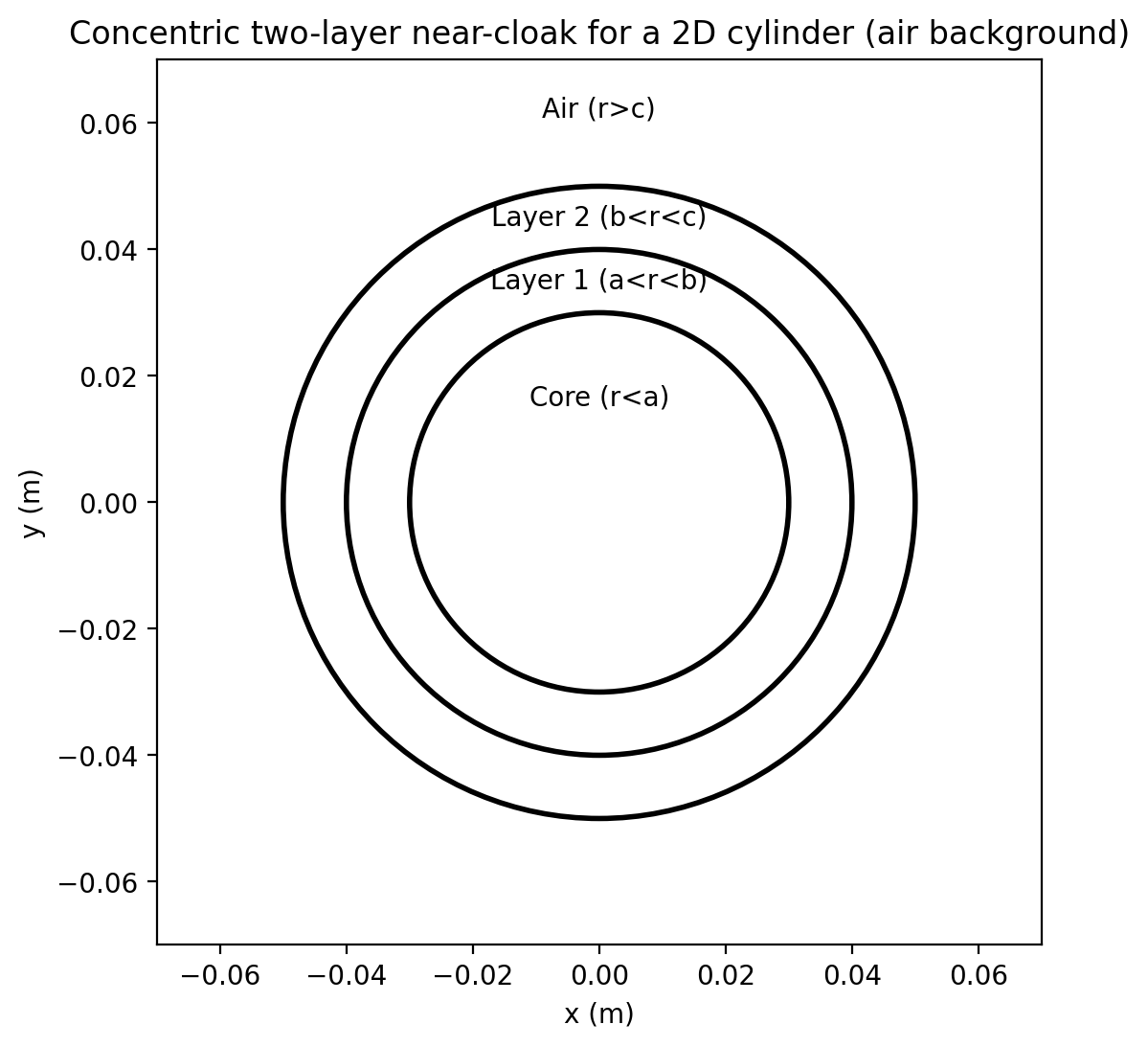}
\caption{Geometry of the concentric two-layer near-cloak for a 2D cylinder in an air background. The core has radius $a$, and the coating layers extend
to radii $b$ and $c$.}
\label{fig:geom}
\end{figure}

We consider time-harmonic acoustics $p(r,\theta)e^{-\ii\omega t}$ in an unbounded air background with density $\rho_0$ and sound speed $c_0$,
so that $k_0=\omega/c_0$. A circular cylinder (core) of radius $a$ is surrounded by two concentric layers of thicknesses $t_1$ and $t_2$,
yielding radii $b=a+t_1$ and $c=a+t_1+t_2$ (Fig.~\ref{fig:geom}). Each region $j\in\{1,2,3\}$ is modeled as an isotropic effective fluid with
density $\rho_j$ and sound speed $c_j$ ($k_j=\omega/c_j$). Within each region the pressure satisfies
\begin{equation}
\nabla^2 p + k_j^2 p = 0.
\end{equation}

We will use dimensionless groups
\begin{equation}
\kappa \equiv k_0 a,\qquad \beta \equiv \frac{b}{a},\qquad \gamma \equiv \frac{c}{a},\qquad
Z_j \equiv \rho_j c_j,\qquad K_j \equiv \rho_j c_j^2,
\end{equation}
and note that low-order multipoles dominate as $k_0c=\gamma\kappa \to 0$.

\subsection{Cylindrical-harmonic expansion}
For an incident plane wave $p_{\mathrm{inc}}(r,\theta)=\exp(\ii k_0 r \cos\theta)$,
\begin{equation}
p_{\mathrm{inc}}=\sum_{m=-\infty}^{\infty} \ii^m J_m(k_0 r)\, e^{\ii m\theta}.
\end{equation}
The scattered field is
\begin{equation}
p_{\mathrm{scat}}=\sum_{m=-\infty}^{\infty} a_m(\omega)\, H_m^{(1)}(k_0 r)\, e^{\ii m\theta},
\end{equation}
with complex coefficients $a_m(\omega)$. In the core we enforce regularity at $r=0$, and in coating layers allow linear combinations of $J_m$ and $Y_m$:
\begin{align}
p_1^{(m)}(r) &= A_{1m} J_m(k_1 r), \\
p_2^{(m)}(r) &= A_{2m} J_m(k_2 r) + B_{2m} Y_m(k_2 r), \\
p_3^{(m)}(r) &= A_{3m} J_m(k_3 r) + B_{3m} Y_m(k_3 r), \\
p_0^{(m)}(r) &= \ii^m J_m(k_0 r) + a_m H_m^{(1)}(k_0 r).
\end{align}

\subsection{Boundary conditions and per-mode linear system}
At $r=a,b,c$ we impose continuity of pressure and radial particle velocity $u_r=-(1/(\ii\omega \rho))\partial_r p$,
equivalently continuity of $(1/\rho)\partial_r p$. For each angular order $m$ this yields a $6\times 6$ linear system
$\mathbf{M}_m(\omega)\bm{u}_m=\bm{b}_m(\omega)$ for $\bm{u}_m=[A_{1m},A_{2m},B_{2m},A_{3m},B_{3m},a_m]^T$.
An explicit construction is provided in App.~\ref{app:linear_system}.

\subsection{Energetic metric: total scattering width}
We use the 2D total scattering width
\begin{equation}
\sigma_{\mathrm{scat}}(\omega)=\frac{4}{k_0}\sum_{m=-\infty}^{\infty} \abs{a_m(\omega)}^2,
\label{eq:sigma}
\end{equation}
computed with symmetric truncation $\abs{m}\le M_{\max}$ chosen for convergence.

\section{Near-cloak design and Fisher-information detectability}
\subsection{Multipole cancellation objective and design procedure}
In the low-frequency regime ($k_0 c \ll 1$), scattering is dominated by the monopole ($m=0$) and dipole ($m=\pm 1$) terms. We therefore define the
design objective
\begin{equation}
\mathcal{J}(\omega_0)\equiv \abs{a_0(\omega_0)}^2 + 2\abs{a_1(\omega_0)}^2,
\label{eq:objective}
\end{equation}
and seek effective-layer parameters that minimize $\mathcal{J}$ at a target frequency $\omega_0$.
In the numerical illustration, we used a constrained random search over $(\rho_2,c_2,\rho_3,c_3)$ with positivity constraints $\rho_j>0$, $c_j>0$
and with bounds selected to avoid trivial singular limits; the reported design is representative rather than claimed globally optimal.

\subsection{Fisher information and CRLB for joint size--material inference}
Assume $K$ far-field samples at angles $\theta_k$ with complex noisy measurements
\begin{equation}
y_k = \mu(\theta_k;\bm{x}) + n_k,
\end{equation}
where $\bm{x}=[a,\rho_1,c_1]^T$ and $n_k$ is circular complex Gaussian noise with variance $\sigma^2$.
We model the scattered far-field angular dependence as
\begin{equation}
\mu(\theta;\bm{x}) \propto \sum_{m=-M}^{M} a_m(\bm{x}) e^{\ii m\theta}.
\label{eq:mu}
\end{equation}
For complex Gaussian noise, a convenient expression for the FIM is
\begin{equation}
I_{ij}(\bm{x})=\frac{2}{\sigma^2}\,\Re\sum_{k=1}^{K}\left(\frac{\partial \mu(\theta_k;\bm{x})}{\partial x_i}\right)^{\!*}
\left(\frac{\partial \mu(\theta_k;\bm{x})}{\partial x_j}\right),
\label{eq:fim}
\end{equation}
which we evaluate using numerical central differences for $\partial\mu/\partial x_i$.
The CRLB satisfies $\mathrm{Cov}(\hat{\bm{x}})\succeq \mathbf{I}(\bm{x})^{-1}$; when $\mathbf{I}$ is ill-conditioned we use the Moore--Penrose pseudoinverse.
To report normalized results without ambiguity in units, we plot ratios such as $\sigma_{\mathrm{bare}}/\sigma_{\mathrm{cloak}}$ and
$\Tr(I_{\mathrm{bare}})/\Tr(I_{\mathrm{cloak}})$ in decibels.

\subsection{Why energetic and information-based invisibility can decouple (low-order analytic view)}
To make the distinction between energetic and inference-based invisibility explicit, consider the low-order truncation $|m|\le 1$ of the far-field model
in Eq.~\eqref{eq:mu}. Using symmetry for a plane-wave excitation of a circularly symmetric structure, the scattered pattern can be written as
\begin{equation}
\mu(\theta;\bm{x}) \approx A_0(\bm{x}) + A_1(\bm{x})\cos\theta + B_1(\bm{x})\sin\theta,
\label{eq:mu_low}
\end{equation}
where $(A_0,A_1,B_1)$ are linear combinations of the complex coefficients $a_0,a_{\pm 1}$. \footnote{For example, one convenient real trigonometric form is
$\mu(\theta)\propto a_0 + a_{1}e^{\ii\theta}+a_{-1}e^{-\ii\theta}$.}
An energetic near-cloak designed by multipole cancellation aims to make $A_0,A_1,B_1$ small at $\omega_0$.
However, the Fisher information depends on \emph{sensitivities} through Eq.~\eqref{eq:fim}. Inserting Eq.~\eqref{eq:mu_low} into Eq.~\eqref{eq:fim} yields
\begin{equation}
I_{ij} \propto \Re\sum_{k}\left(\partial_{x_i}A_0 + \partial_{x_i}A_1\cos\theta_k + \partial_{x_i}B_1\sin\theta_k\right)^{\!*}
\left(\partial_{x_j}A_0 + \partial_{x_j}A_1\cos\theta_k + \partial_{x_j}B_1\sin\theta_k\right),
\label{eq:fim_low}
\end{equation}
so that $I_{ij}$ can remain large even when $A_0,A_1,B_1$ themselves are suppressed, provided the derivatives
$\partial_{x}A_0,\partial_{x}A_1,\partial_{x}B_1$ do not vanish. This explains, at a model level, why a coating can yield large reductions in
$\sigma_{\mathrm{scat}}$ while reducing $\Tr(\mathrm{FIM})$ only modestly: multipole cancellation constrains \emph{field amplitudes} at $\omega_0$, whereas
information suppression requires simultaneous reduction of \emph{parameter sensitivities}. The latter is a stronger (and generally multi-objective) requirement.

\section{Results: air-background numerical illustration}
We consider an air background with $\rho_0=\SI{1.21}{kg.m^{-3}}$ and $c_0=\SI{343}{m.s^{-1}}$.
The core is modeled as an effective fluid approximating a high-impedance object with $\rho_1=\SI{1200}{kg.m^{-3}}$ and $c_1=\SI{2500}{m.s^{-1}}$.
Geometry: $a=\SI{3}{cm}$, $t_1=\SI{1}{cm}$, $t_2=\SI{1}{cm}$ (so $b=\SI{4}{cm}$ and $c=\SI{5}{cm}$). A representative two-layer design uses
effective parameters (layer 1) $\rho_2=\SI{8.60}{kg.m^{-3}}$, $c_2=\SI{387}{m.s^{-1}}$ and
(layer 2) $\rho_3=\SI{0.255}{kg.m^{-3}}$, $c_3=\SI{455}{m.s^{-1}}$, targeting $f_0=\omega_0/(2\pi)=\SI{500}{Hz}$.
For FIM/CRLB curves we use $K=60$ angles, truncation $\abs{m}\le 6$, and complex Gaussian noise standard deviation $\sigma=10^{-3}$ (arbitrary units).

\begin{figure}[t]
\centering
\includegraphics[width=\columnwidth]{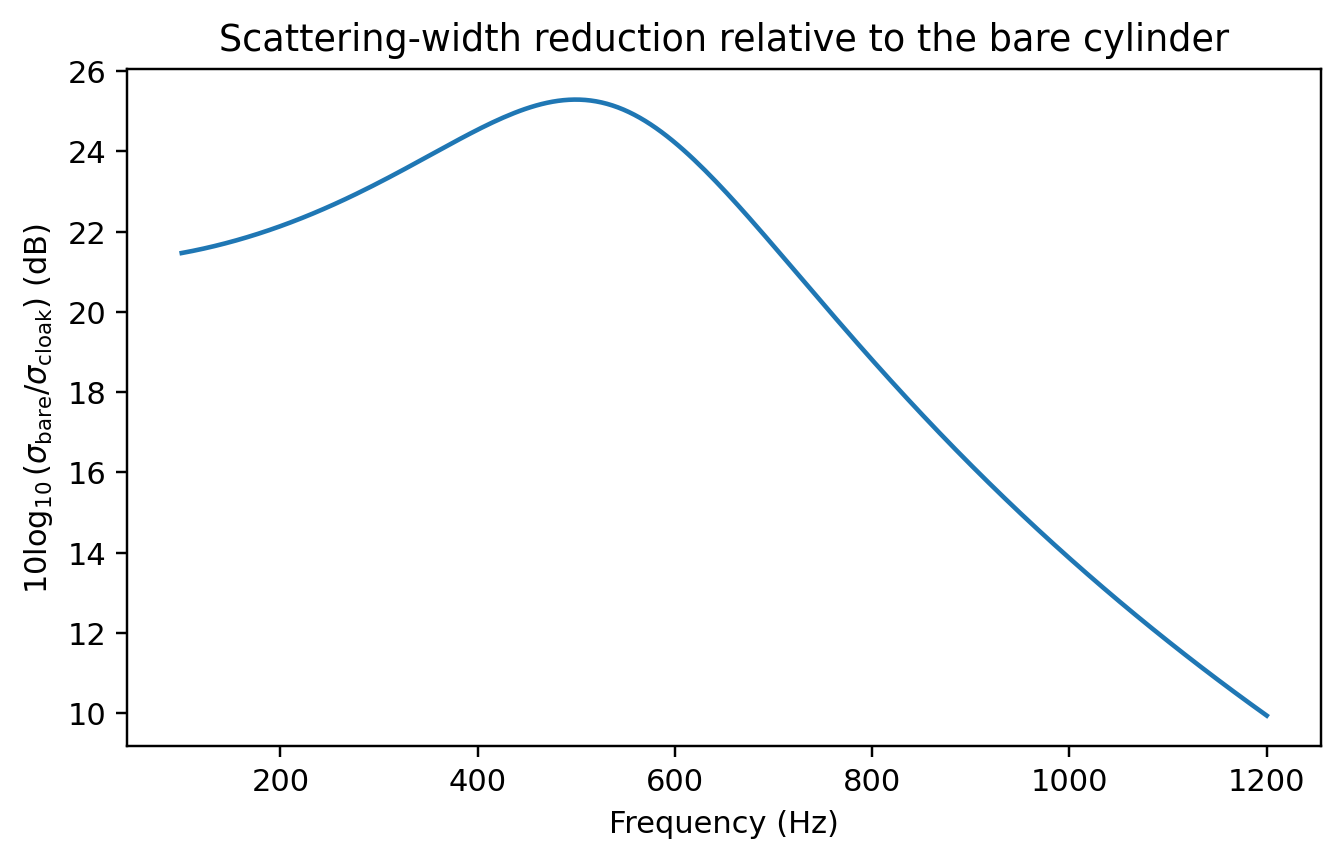}
\caption{Normalized scattering-width reduction relative to the bare cylinder, $10\log_{10}(\sigma_{\mathrm{bare}}/\sigma_{\mathrm{cloak}})$, versus frequency.
The near-cloak is designed near $f_0\approx \SI{500}{Hz}$ and achieves a peak reduction of $\sim 25$~dB in this illustration.}
\label{fig:scatred}
\end{figure}

\begin{figure}[t]
\centering
\includegraphics[width=\columnwidth]{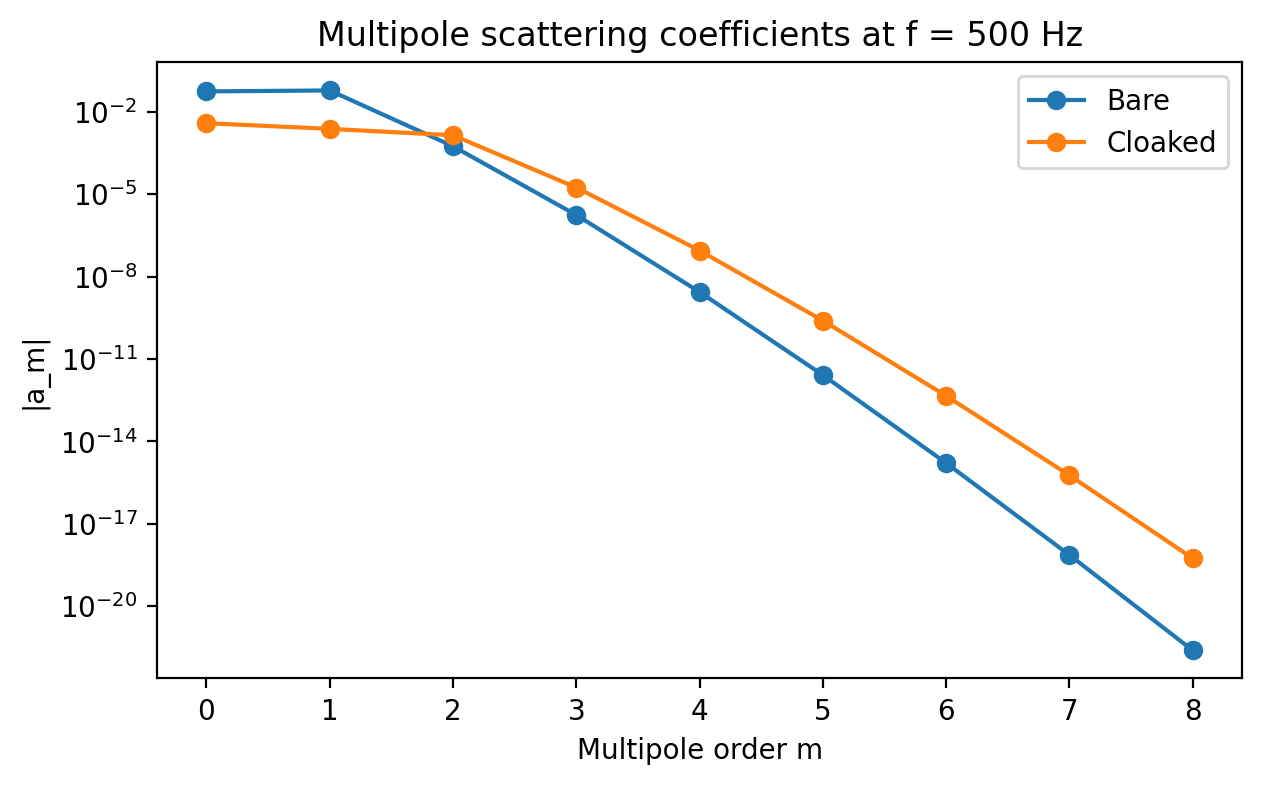}
\caption{Magnitudes of multipole scattering coefficients $\abs{a_m}$ at $f_0$ for bare and cloaked configurations. The coating suppresses the dominant
monopole ($m=0$) and dipole ($m=1$) coefficients.}
\label{fig:multipoles}
\end{figure}

\begin{figure}[t]
\centering
\includegraphics[width=\columnwidth]{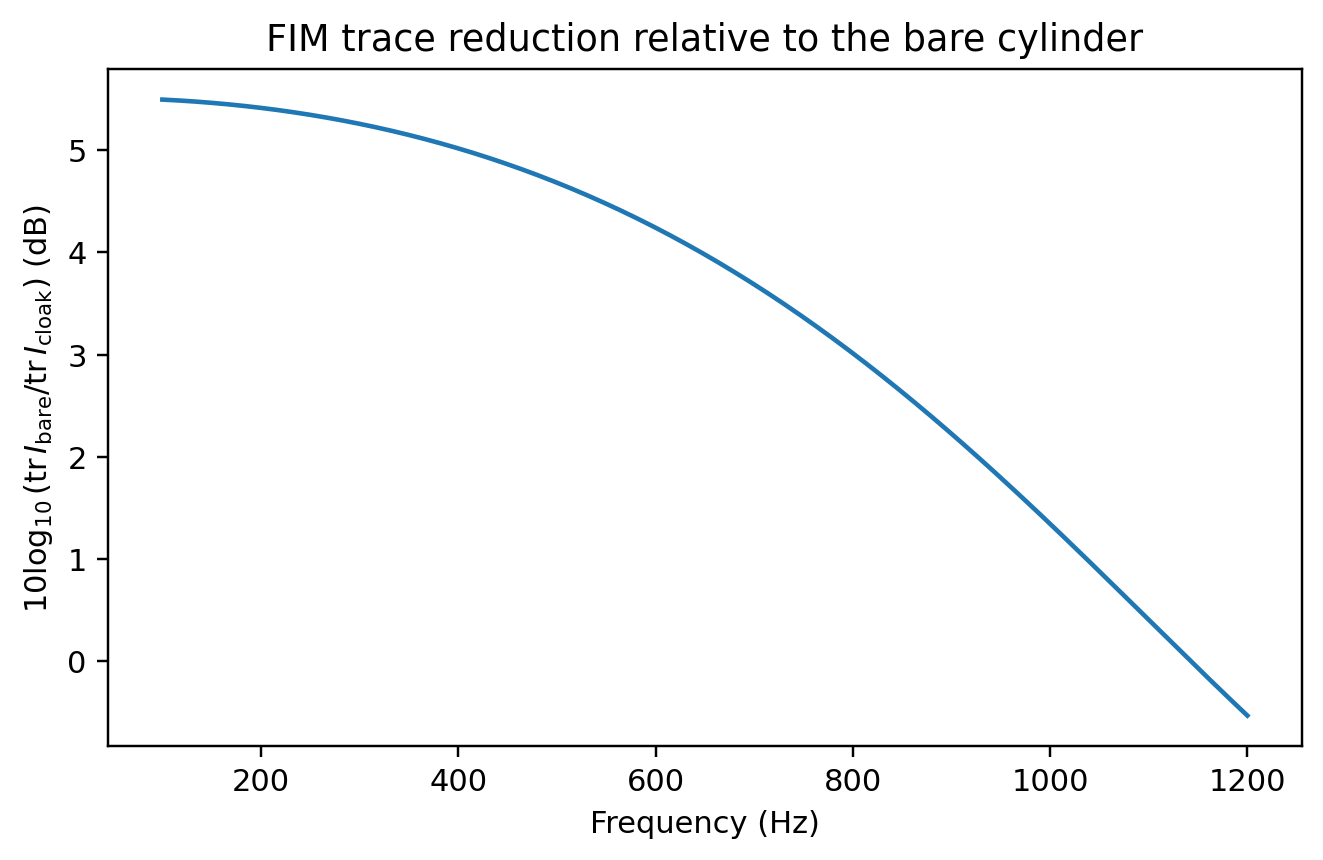}
\caption{Normalized information-based reduction relative to the bare cylinder, shown as $10\log_{10}(\Tr I_{\mathrm{bare}}/\Tr I_{\mathrm{cloak}})$ for joint
estimation of $\bm{x}=[a,\rho_1,c_1]^T$ from noisy far-field data.}
\label{fig:fimred}
\end{figure}

\begin{figure}[t]
\centering
\includegraphics[width=\columnwidth]{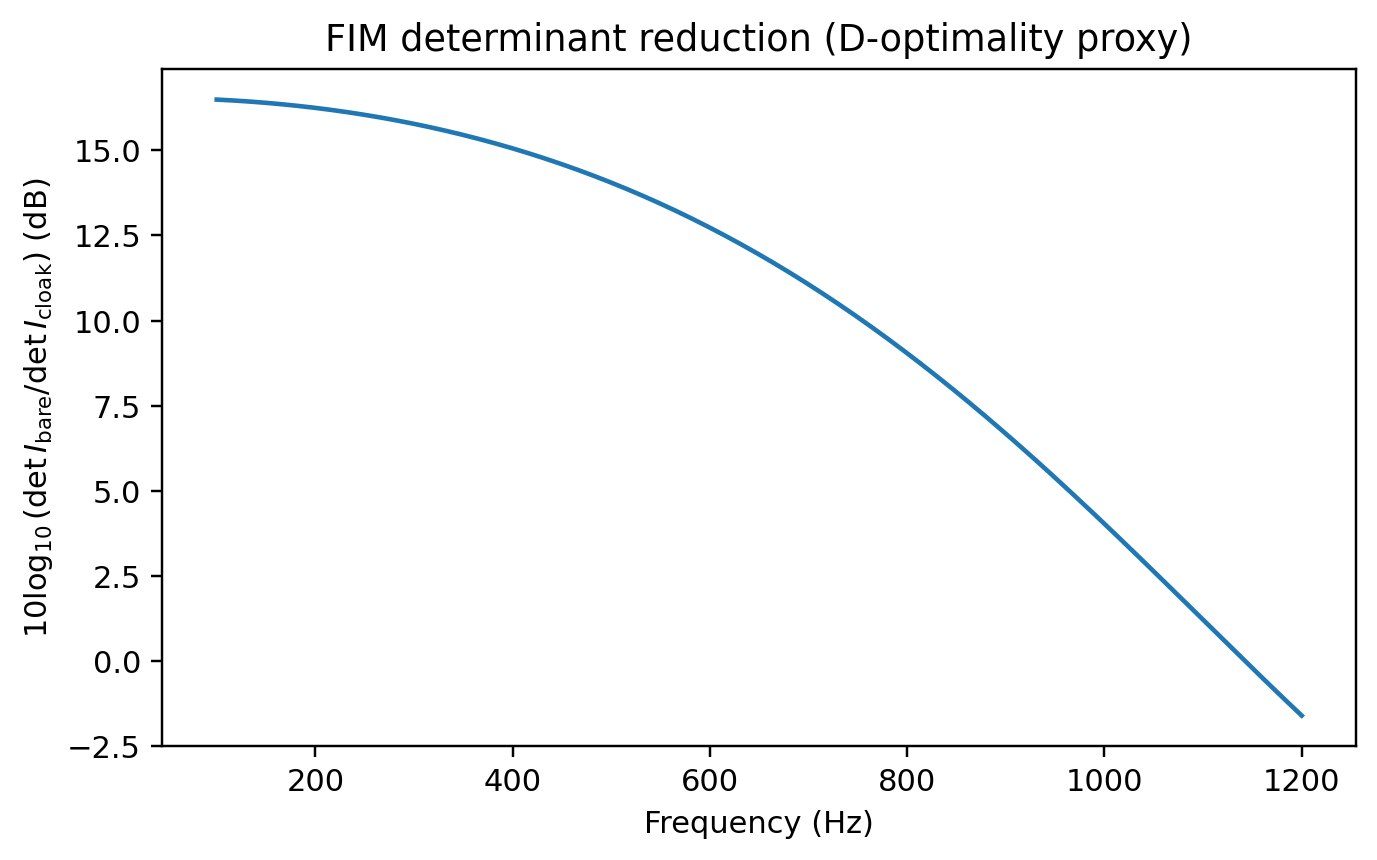}
\caption{Determinant-based Fisher-information reduction, $10\log_{10}(\det I_{\mathrm{bare}}/\det I_{\mathrm{cloak}})$, providing a D-optimality proxy that
emphasizes worst-direction identifiability.}
\label{fig:logdet}
\end{figure}

\begin{figure}[t]
\centering
\includegraphics[width=\columnwidth]{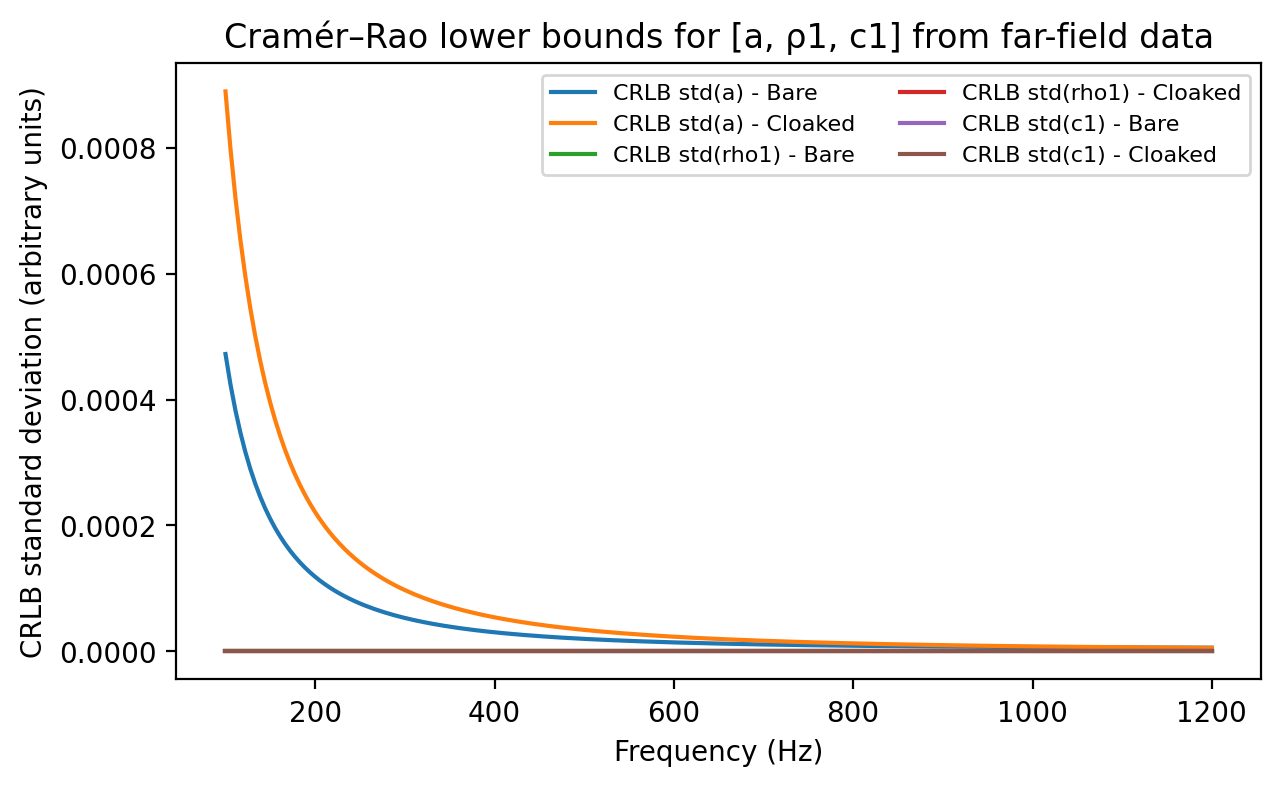}
\caption{Cram\'er--Rao lower bounds (CRLB) for the standard deviations of $a$, $\rho_1$, and $c_1$ inferred from far-field data. The near-cloak generally
increases CRLBs near the design frequency, indicating reduced parameter identifiability.}
\label{fig:crlb}
\end{figure}

\begin{figure}[t]
\centering
\includegraphics[width=\columnwidth]{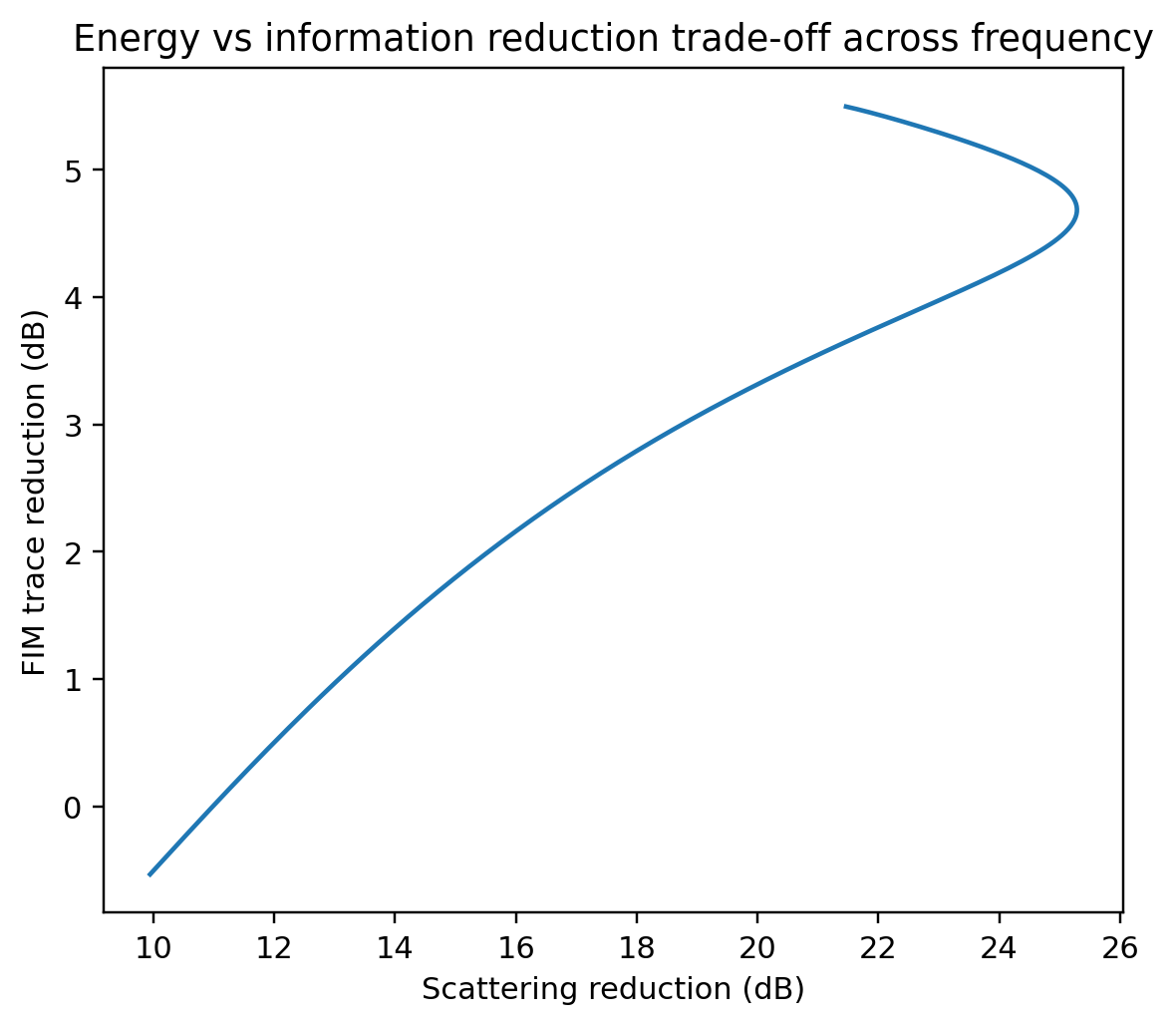}
\caption{Trade-off between energetic scattering reduction and information reduction across frequency, using the normalized metrics shown in
Figs.~\ref{fig:scatred} and~\ref{fig:fimred}. Different frequencies correspond to different operating points on the curve.}
\label{fig:tradeoff}
\end{figure}

Figure~\ref{fig:scatred} shows the scattering-width reduction $10\log_{10}(\sigma_{\mathrm{bare}}/\sigma_{\mathrm{cloak}})$ versus frequency, with a peak
reduction $\sim 25$~dB near the design frequency. Figure~\ref{fig:multipoles} shows that the monopole and dipole coefficients are strongly suppressed at $f_0$.
Figure~\ref{fig:fimred} reports the normalized FIM-trace reduction $10\log_{10}(\Tr I_{\mathrm{bare}}/\Tr I_{\mathrm{cloak}})$, while Fig.~\ref{fig:logdet}
reports the determinant-based proxy $10\log_{10}(\det I_{\mathrm{bare}}/\det I_{\mathrm{cloak}})$ (D-optimality). Figure~\ref{fig:crlb} reports CRLB
standard deviations for $[a,\rho_1,c_1]$. Finally, Fig.~\ref{fig:tradeoff} summarizes the frequency-dependent trade-off between scattering suppression and
information suppression.

\begin{figure}[t]
\centering
\includegraphics[width=\columnwidth]{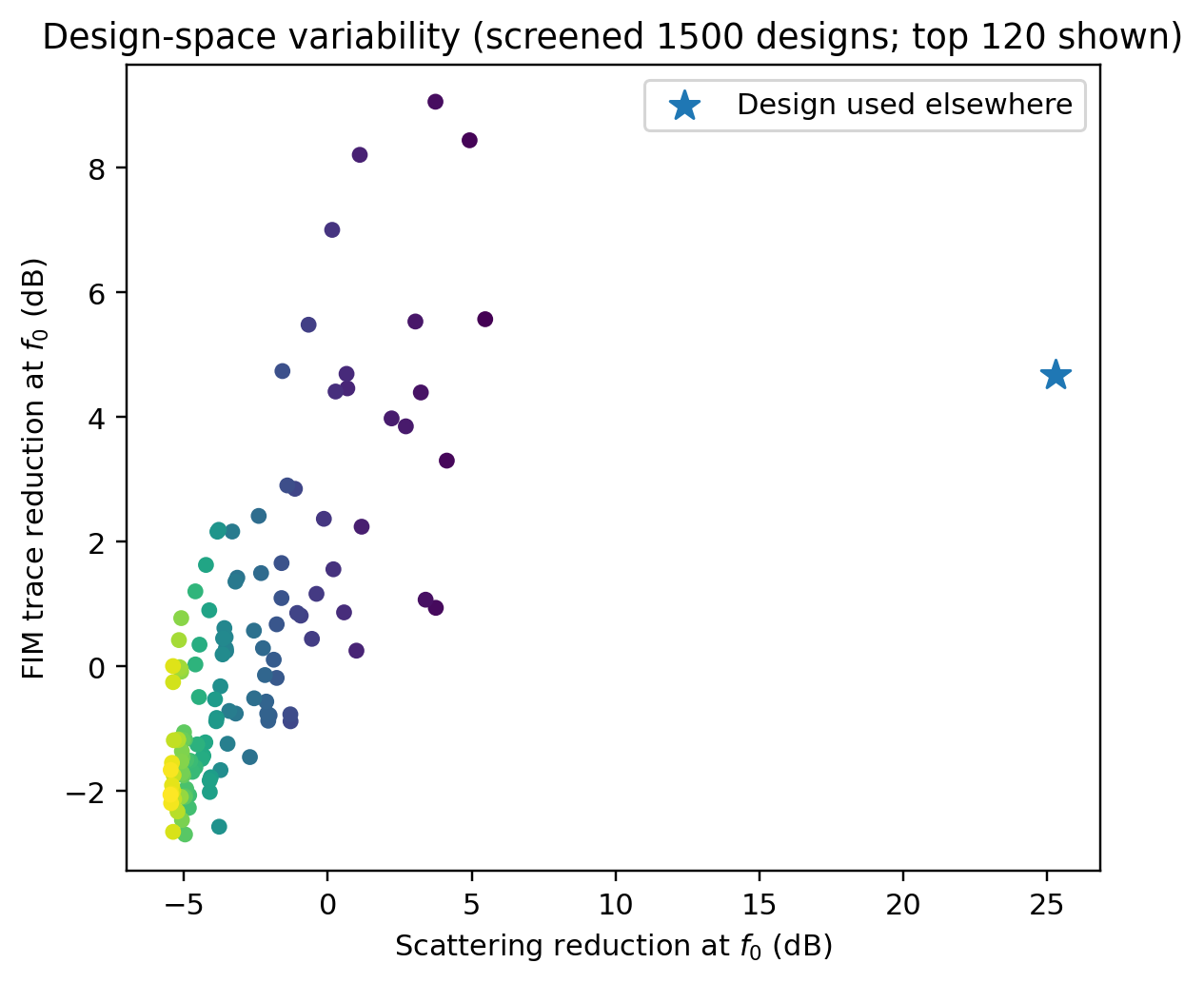}
\caption{Design-space variability at the design frequency. We screened $N=1500$ random two-layer effective-fluid designs under positivity constraints,
ranked them by the low-order multipole objective in Eq.~\eqref{eq:objective}, and computed energetic and information-based reductions for a subset of 120 screened designs (those with the smallest objective value).
Points show the joint behavior of scattering reduction and Fisher-information reduction at $f_0$. The star marks the design used in
Figs.~\ref{fig:scatred}--\ref{fig:tradeoff}.}
\label{fig:designspace}
\end{figure}

\begin{figure}[t]
\centering
\includegraphics[width=\columnwidth]{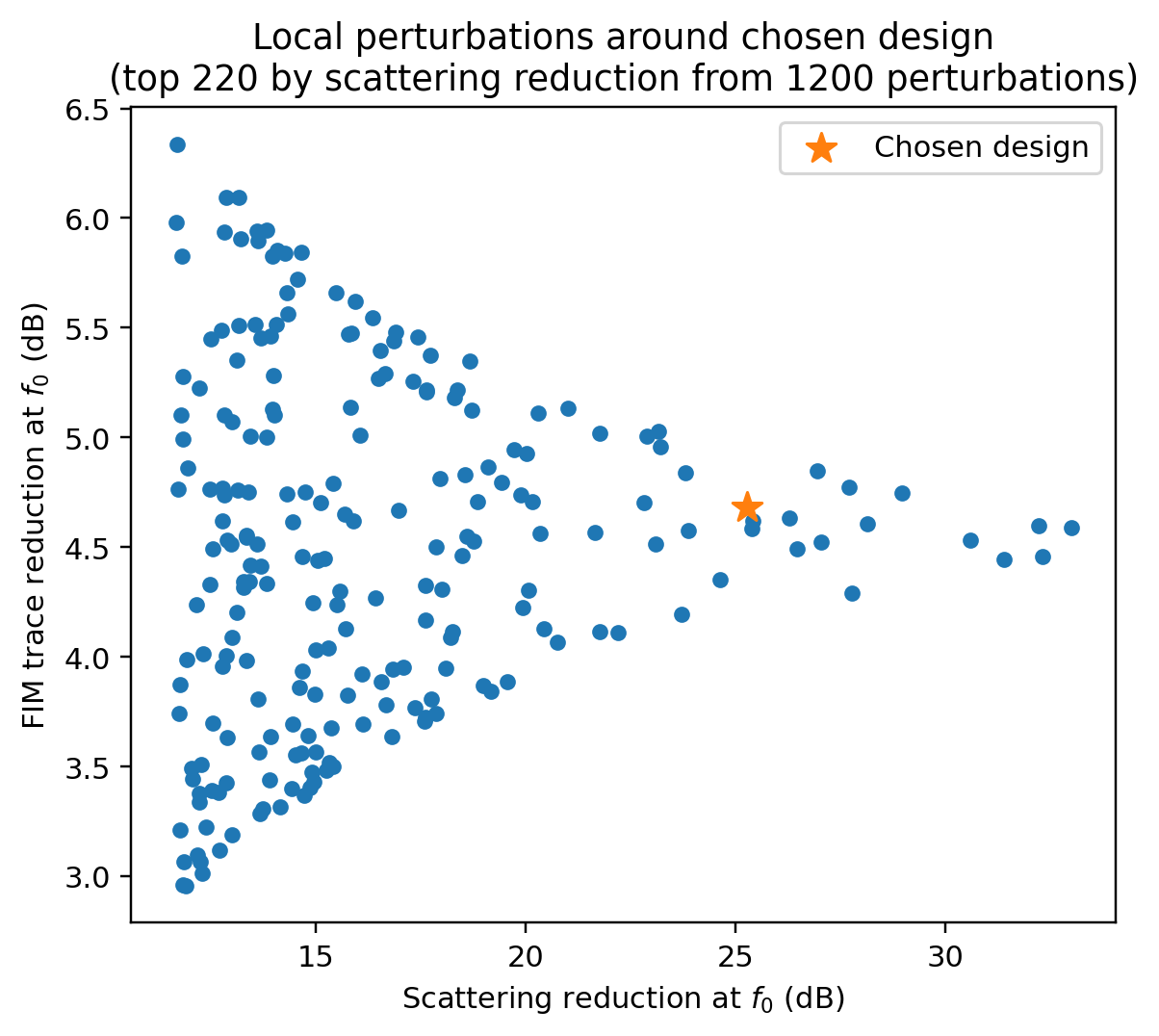}
\caption{Local robustness and trade-off near the chosen design. We generated random multiplicative perturbations of the layer parameters around the chosen
design and retained the best 220 perturbations by scattering reduction at $f_0$. The resulting cloud shows that sizable energetic suppression can be maintained
while Fisher-information reduction remains comparatively modest, consistent with the analytic decoupling argument in Sec.~III.C.}
\label{fig:local}
\end{figure}

In addition to reporting frequency-dependent metrics for a single near-cloak, we quantify \emph{generality} in two complementary ways.
First, we perform a screened random search over effective-layer parameters and evaluate the joint behavior of energetic suppression and
information suppression at $f_0$ (Fig.~\ref{fig:designspace}). Second, we probe local robustness by perturbing the chosen design parameters and reporting
the resulting trade-off cloud (Fig.~\ref{fig:local}). For these design-space diagnostics we use a reduced FIM model (24 angles, $|m|\le 3$) to keep the
computational cost tractable, while retaining the same parameter vector $\bm{x}=[a,\rho_1,c_1]^T$.

\section{Discussion}
The results highlight a PRE-relevant distinction between energetic and inferential notions of invisibility.
While $\sigma_{\mathrm{scat}}$ summarizes scattered-field energy [Eq.~\eqref{eq:sigma}], the FIM depends on parameter sensitivities through derivatives
of $\mu(\theta;\bm{x})$ [Eq.~\eqref{eq:mu}] as made explicit by Eq.~\eqref{eq:fim}. Consequently, minimizing $\sigma_{\mathrm{scat}}$ at $\omega_0$ does not
generally minimize $\Tr(\mathrm{FIM})$ or $\log\det(\mathrm{FIM})$, and may lead to markedly different behavior across frequency.

The narrowband character of the near-cloak and its degradation away from $\omega_0$ are consistent with passivity/causality and delay--bandwidth limitations
for passive cloaks~\cite{MonticoneAlu2016,HashemiZhang2010}. The Fisher-information viewpoint provides a task-aware way to quantify identifiability and
conditioning, and suggests multi-objective designs that suppress both scattering and Fisher information under physical constraints.

Finally, we emphasize scope: we adopt an effective-fluid model for both layers and the core to retain a closed-form modal structure. Extensions to elastic cores
or coatings are feasible but require elastodynamic potentials and traction boundary conditions, substantially increasing algebraic complexity.

\section{Conclusions}
We presented a fully theoretical framework for two-layer multipole-cancellation near-cloaking of a 2D cylinder in air and introduced Fisher-information and
Cram\'er--Rao metrics to quantify inference-based detectability for joint size--material estimation. Normalized results demonstrate that large reductions in total
scattering can correspond to smaller reductions in FIM-based detectability, emphasizing that invisibility depends on the observer's task. This motivates future
multi-objective theoretical designs constrained by physical bounds on passive cloaks.

\appendix
\section{Explicit per-mode linear system}
\label{app:linear_system}
For a fixed angular order $m$, define unknown coefficients
$\bm{u}_m=[A_{1m},A_{2m},B_{2m},A_{3m},B_{3m},a_m]^T$.
Continuity of $p$ and $(1/\rho)\partial_r p$ at $r=a,b,c$ yields $\mathbf{M}_m(\omega)\bm{u}_m=\bm{b}_m(\omega)$ with
\begin{widetext}
\small
\begin{equation}
\mathbf{M}_m=
\begin{pmatrix}
J_m(k_1 a) & -J_m(k_2 a) & -Y_m(k_2 a) & 0 & 0 & 0\\
\frac{k_1}{\rho_1}J_m'(k_1 a) & -\frac{k_2}{\rho_2}J_m'(k_2 a) & -\frac{k_2}{\rho_2}Y_m'(k_2 a) & 0 & 0 & 0\\
0 & J_m(k_2 b) & Y_m(k_2 b) & -J_m(k_3 b) & -Y_m(k_3 b) & 0\\
0 & \frac{k_2}{\rho_2}J_m'(k_2 b) & \frac{k_2}{\rho_2}Y_m'(k_2 b) & -\frac{k_3}{\rho_3}J_m'(k_3 b) & -\frac{k_3}{\rho_3}Y_m'(k_3 b) & 0\\
0 & 0 & 0 & J_m(k_3 c) & Y_m(k_3 c) & -H_m^{(1)}(k_0 c)\\
0 & 0 & 0 & \frac{k_3}{\rho_3}J_m'(k_3 c) & \frac{k_3}{\rho_3}Y_m'(k_3 c) & -\frac{k_0}{\rho_0}{H_m^{(1)}}'(k_0 c)
\end{pmatrix},
\end{equation}

\begin{equation}
\bm{b}_m=
\begin{pmatrix}
0\\
0\\
0\\
0\\
\ii^m J_m(k_0 c)\\
\ii^m \frac{k_0}{\rho_0}J_m'(k_0 c)
\end{pmatrix}.
\end{equation}
\normalsize
\end{widetext}
Here $J_m',Y_m',{H_m^{(1)}}'$ denote derivatives with respect to their arguments.
Solving for $\bm{u}_m$ yields $a_m$ as the sixth component.

\bibliography{refs}

\end{document}